\documentclass[12pt,preprint]{aastex}
\usepackage{natbib}
\usepackage{amssymb, amsmath, amsbsy, epsfig, epsf, slashbox, rotating}
\usepackage{color}

\slugcomment{Accepted for publication in {\it The Astrophysical Journal Letters}}
\begin{document}

\title{Constraining Very High Mass Population III Stars through He II Emission in Galaxy BDF-521 at z = 7.01}
\author{Zheng Cai \altaffilmark{1,2}, Xiaohui Fan \altaffilmark{1}, Linhua Jiang\altaffilmark{3,7}, Romeel Dav\'{e} \altaffilmark{1,4}, S. Peng Oh \altaffilmark{5}, Yujin Yang\altaffilmark{6}, Ann Zabludoff \altaffilmark{1}}
\affil{Steward Observatory, University of Arizona, Tucson, AZ 85721, USA}
\affil{Physics Department, University of Arizona, Tucson, AZ 85721, USA}
\affil{School of Earth and Space Exploration, Arizona State University, Tempe, AZ, 85287-1504, USA}
\affil{University of the Western Cape, 7535 Bellville, Cape Town, South Africa}
\affil{Dept. of Physics Broida Hall University of California Santa Barbara, CA 93106-9530, USA}
\affil{Argelander-Institut fuer Astronomie, Auf dem Huegel 71 Bonn, 53121 Germany}
\affil{Hubble Fellow}
%\date{\today}
%\maketitle

\altaffiltext{1} {Email: caiz at email.arizona.edu}
% Usually omit these for ApJ or MNRAS style files:
%\tableofcontents
%
%\listoffigures
%
%\listoftables

\begin{abstract}
Numerous theoretical models have long proposed that a strong He II $\lambda1640$ emission line 
is the most prominent and unique feature of massive Population III (Pop III) stars in high redshift galaxies. 
The He II $\lambda 1640$ line strength can constrain the mass and IMF 
of Pop III stars. 
We use F132N narrowband filter on the Hubble Space Telescope's (HST) Wide Field Camera 3 (WFC3) 
to look for strong He II $\lambda 1640$ emission in the galaxy BDF-521 at $z=7.01$, one of the most distant spectroscopically-confirmed galaxies to date. 
Using deep F132N narrowband imaging, together with
our broadband imaging with F125W and F160W filters,
we do not detect He II emission from this galaxy, but place a 2$\sigma$ upper limit on the flux of  $5.3\times10^{-19}\ \rm{ergs}\ \rm{s}^{-1}\ \rm{cm}^{-2}$. This measurement
corresponds to a 2$\sigma$ upper limit on the Pop III star formation rate (SFR$_{\rm{PopIII}}$) of
$\sim 0.2\ \rm{M}_{\odot}\ \rm{yr}^{-1}$, assuming a Salpeter IMF
with $50\lesssim M/M_{\odot}\lesssim 1000$. From the high signal-to-noise 
broadband measurements in F125W and F160W, we fit the UV continuum for BDF-521. 
The spectral flux density is $\sim 3.6\times10^{-11} \times \lambda^{-2.32}\ \rm{ergs}\ \rm{s}^{-1}\
\rm{cm}^{-2}$\ \AA$^{-1}$, which corresponds to an overall unobscured SFR of $\sim$ 5 M$_{\odot}\ \rm{yr}^{-1}$. 
Our upper limit on SFR$_{\rm{PopIII}}$ suggests that massive Pop III stars represent
$\lesssim 4$\% of the total star formation.
Further, the HST high resolution imaging suggests that BDF-521 
is an extremely compact galaxy, with a half-light radius of $0.6$ kpc.
\end{abstract}

%Section heading
\section{Introduction}
%% First paragraph: general importance of Pop III stars. 
The first generation stars, i.e., Population III (Pop III) stars with zero metallicity, form out of primordial material left over from the Big Bang. They are believed to produce a significant amount of energy in the ultraviolet (UV) and to serve as the main sources to re-ionize the inter-galactic medium (IGM). %The remnant black holes of Pop III stars seeded the growth of super-massive black holes (SMBHs) at $z=6-7$ we observed today \cite[e.g.][]{mortlock11}.
 The confirmation and analysis of Pop III stars or Pop III hosting high-redshift galaxies are among the most crucial goals to studies of the origin and evolution of galaxies in the early Universe.  

In recent years, a number of galaxy candidates at $z=7-10$ have been found, mainly due to ambitious near-infrared (NIR) imaging surveys with greater depths and larger fields of view \cite[e.g.][]{yan12, robertson13}. With the biggest ground-based telescopes, significant progress has also been made in spectroscopically confirming Ly$\alpha$ emission from galaxies at $z\sim 7$ 
\citep{iye06, vanzella11, ono12, schenker12, finkelstein13}. 
However, the fraction of galaxy candidates at $z=7$ that have Ly$\alpha$ emission is still very small, mainly because a rising IGM neutral fraction to suppress the Ly$\alpha$ emission 
\cite[e.g.][]{caruana14, bolton13, fontana10}.
On the other hand, sensitive mm/sub-mm facilities like the Atacama Large Millimeter Array (ALMA) have allowed CO and [CII] to 
serve as possible tracers of metallicity and to address the optical bias in star formation rate (SFR) measurements for galaxies at $z\sim 7$ \cite[e.g.][]{ouchi13}. 

Direct detections of Pop III signatures in high redshift galaxies, however, remain a major challenge \cite[e.g.][]{cai11}. Theoretical investigations suggest that Pop III stars have a high characteristic mass, 
and short life spans on the order of a few million years \cite[e.g.][]{bromm11}.
Feedback from the first stars will metal-enrich the local primordial interstellar medium (ISM) and quench further Pop III star formation \citep{yoshida04, muratov13}. Therefore, Pop III stars should only be dominant in very young galaxies at high redshift. On average, younger galaxies tend to be less massive than more mature systems at given redshift, and thus they are intrinsically fainter in the continuum bands, making the direct detection very challenging with today's telescopes. \citet{salvaterra11} suggest that, at $z\gtrsim 6$, Pop-III dominated ($>50\%$ of total luminosity) galaxies are extremely faint ($M_{\rm{UV}}>-14$), which is below current detection limits. 

The He II $\lambda$1640 line from high redshift galaxies has been long suggested to depend strongly on both the IMF and the stellar metallicity \cite[e.g.][]{schaerer02}. 
Hard ionizing radiation, particularly photons that ionize Helium, is greatly enhanced by the high surface temperature of Pop III stars. Therefore, the resulting strong recombined He II $\lambda1640$ line is a direct footprint of Pop III stars.  Moreover, the He II $\lambda$1640 line is much less complicated to model and interpret, compared to the Ly$\alpha$ recombination line \citep{schaerer03}. Also, He II $\lambda1640$ photons pass more easily through the partially neutral IGM at high redshift than Ly$\alpha$ photons \citep{tumlinson00}. Using the model of \citet{schaerer02}, one expects to detect strong He II $\lambda$1640 emission in $z\sim7$ spectroscopically-comfirmed galaxies if $30\%$ of the SFR is contributed by Pop III stars with a top-heavy IMF. 
%Therefore, many efforts have been put trying 

In the past few years, a few observations have probed Pop III signatures in high-redshift galaxies by looking for strong He II emission \citep{dawson04, nagao05, nagao08, cai11}. These studies have reported nondetections, constraining the Pop III star formation rate (hereafter SFR$_{\rm{PopIII}}$) to less than a few M$_{\odot}$ yr$^{-1}$. Note the narrow He II line (FWHM $\sim$ a few hundred km s$^{-1}$) is one of the Pop III star signatures, which is distinguished from broad He II emission from AGN \citep{eldridge12}. The line width measurement through spectroscopy is required to unambiguously confirm the He II from Pop III stars.
\citet{cassata13} find that $\sim$ 10\% of galaxies at $z\sim 3$ with i$_{\rm{AB}} < 24.75$ show He II emission, with rest-frame equivalent widths (EW)  $\sim$ $1-7$ \AA, and argue that some of the galaxies with narrow He II lines are compatible with the predictions of a Pop III scenario. However, the EWs of these He II emission lines are not exceptionally high compared with those from some other astrophysical objects, so the He II emission also might be interpreted as metal-poor nebular He II emission. 

%The unique signature of 
% \cite[e.g.][]{skillman93}. 

More metal-poor galactic environments could be expected at higher redshifts. %Recent simulation shows that on average, the fraction of Pop III stars drop from 0.19 at $z=10$ to 0.07 $z=5$ \citep{salvaterra11}. 
It is thus important to search for secure Pop III signatures in the highest redshift galaxies. At z $\gtrsim$ 7, the He II $\lambda1640$ line is redshifted into the J-band. Strong near-infrared sky lines make it challenging to conduct ground-based deep observations. The Wide Field Camera 3 (WFC3) on HST has a high throughput, and does not suffer a strong IR background from the atmosphere. Therefore, it is the ideal tool for studying rest-frame ultraviolet spectral features for galaxies at $z\gtrsim7$. 

 BDF-521 is one of the faintest spectroscopically-confirmed galaxies at $z\sim 7$ \citep{vanzella11}. Moderate S/N ground-based observations suggest that it could have the bluest UV slope (Vanzella et al. 2011). These results suggest that BDF-521 is among the youngest galaxies with low metallicity at $z\sim 7$. He II $\lambda 1640$ for this galaxy lies within the WFC3/F132N narrowband filter.
 In this paper, we report a non-detection of the He II $\lambda1640$ emission in this galaxy, which enables us to put the most stringent upper limit to date on massive Pop-III star formation in galaxies at the end of reionization epoch. In addition, with the HST broadband imaging, we measure the UV continuum level, UV slope, and galaxy morphology. %Th important complementary dataset to ground-based observations with only a moderate S/N. 
This paper is organized as follows: 
In \S2, we discuss our observing strategies and data reduction. In \S3, we provide the upper limit of He II $\lambda 1640$ line flux, the measurements of UV continuum level, slope, and galaxy morphology. In \S4, we discuss the implications of our results on the formation of Pop III stars. Throughout this paper, we adopt a conventional cosmology: $\Omega_\Lambda$ = 0.7, $\Omega_{m}$ = 0.3,  and H$_0$ = 70 km s$^{-1}$ Mpc$^{-1}$.

%% Second paragraph: high-z; far-IR, submm observations on high-z QSO galaxy relation
%% Nearby AGN host galaxy in UV, near-IR, the difficulty of observing far-UV QSO host

%% Previous work using DLAs as a coronagraph 

%The far-UV emission directly trace the young and massive SFR in the QSO host galaxies.  

\section{{\it HST} Observations of Galaxy BDF-521 and Data Reduction}

The galaxy BDF-521 is spectroscopically-confirmed by \citet{vanzella11} to be at $z=7.008\pm0.002$. This galaxy is selected from the imaging survey obtained
with VLT/Hawk-I \citep{castellano10b}. Spectroscopy shows a 
clear, asymmetric Ly$\alpha$ emission line with a total luminosity of $7.1\pm0.7\times 10^{42}$ erg s$^{-1}$, corresponding to an overall SFR$_{\rm{Ly\alpha}}$ of $6.5\pm0.7$ M$_\odot$ yr$^{-1}$ (uncorrected for Gunn-Peterson absorption, which reduces Ly$\alpha$ flux). %It is one of the highest redshift galaxies known to date with a secure spectroscopic redshift. 
At this redshift, He II $\lambda1640$ falls at 13133\AA , fully in the sensitive part 
of the WFC3/IR F132N narrowband filter ($\lambda_c=13188$ \AA\ and FWHM$=162$ \AA), allowing us to use HST narrowband imaging to search for HeII emission. %BDF-521 was observed by HST WFC3 in June 2012.  
11 orbits ($\sim$ a 30,800 sec of integration) are allocated in F132N to probe the He II emission. 
A 4-orbit integration were allocated to measure 
the continuum through the F125W and F160W filters, with 2 orbits for each broadband filter. The broadband imaging is 
used to perform an accurate continuum subtraction in the narrowband, as well as to measure the unobscured SFR and galaxy morphology. Our observations are designed to detect He II emission at the $3\sigma$ level if $30\%$ of the star formation in the galaxy is contributed by Pop III stars with a Salpeter IMF with 50 $\lesssim$ M/M$_\odot$ $\lesssim$ 500. 

%why 11-orbit ??? state this after writing the discussions part... 

The entire set of observations consists of four visits. 
The F132N narrowband imaging is separated into three individual visits. The first and second 
visits both contain four orbits. The third visit contains three orbits. Each orbit is slightly dithered relative to previous orbit. The dithering offset between each orbit is determined by standard WFC3/IR 4-point 
dither sequence. We conduct the broadband imaging in the fourth visit. In this visit, both the F125W and F160W observations 
contain two orbits, with each orbit divided into a 4-point dither sequence. This broadband observational design is the 
same with that in Cai et al. (2011).

All the science images were reduced by the WFC3/IR Multidrizzle pipelines \citep{koekemoer02}. 
In each filter, the images were combined with an output scale = 0.06$''$, 0.48 of the linear size of the original pixel size. We simultaneously shrink the area of 
the input pixels  by adjusting the pixfrac parameter to 0.7 (also see Cai et al. 2011). 
High resolution images from the F132N, F125W, and F160W bands are shown in Figure 1.

\section{Galaxy Properties and Upper Limit of He II $\lambda1640$ Flux}
%\subsection{Photometry and }
Following the method described in \citet{cai11}, we perform the photometry using SExtractor \citep{bertin96} using the root mean square map generated by Multidrizzle \citep{cai11}. %(Casertano et al. 2000).
The photometry is measured for BDF-521 in all three filters (F125W, F160W, and F132N) using a circular aperture with a diameter of 0.45''. The aperture is determined by the F132N image ($2.5\times$ half-light radius). 
%We measure the fluxes in the broadband (F125W and F160W) and the narrowband (F132N) images using the same circular aperture with the diameter of 0.48'' determined from the F132N image. 
 BDF-521 has an absolute magnitude $M_{\rm{AB}} (1500 \rm{\AA}) = -19.8$, comparable to $0.4\times L^*_{z=3}$. The photometry results are listed in Table 1. 
A standard power-law continuum is assumed with $f_{\rm{con}} = A (\lambda/1 \rm{\AA})^\beta$, where A is a constant. %and a He II emission line flux $F_{\rm{HeII}}$ is fit using a Guassian function \citep{cai11}. 
The spectral energy distribution (SED) is then fitted by the power-law continuum based on broadband photometry in F125W and F160W (Figure 2). We find

\begin{equation}
\label{eq:spectra}
f_{con}(\lambda) = ( 3.60\pm 0.22 ) \times 10^{-11} (\lambda/1\ \rm{\AA})^{-2.32\pm0.43}.  
\end{equation}
After subtracting the continuum contribution in the narrowband image, we obtain the line flux of He II ($F_{\rm{HeII}}$):  
\begin{equation}
F_{\rm{HeII}}= (-0.7 \pm 0.5) \times 10^{-18}\ \rm{ergs}\ \rm{s}^{-1}\ \rm{cm}^{-2}, 
\end{equation}
i.e., a non-detection.  2-$\sigma$ upper limit of He II flux is $F_{\rm{HeII}}\le 1.0\times 10^{-18}$ ergs s$^{-1}$ cm$^{-2}$, which is the most stringent constraint to date for $z\sim 7$ galaxies. 

 At $z=7.01$, this corresponds to a total HeII luminosity of 
$L_{\rm{HeII}}= F_{\rm{HeII}}\times 4\pi D_L^2\le (-4.1$ $\pm 2.9)\times 10^{41}\ \rm{ergs}\ \rm{s}^{-1}$, 
or a non-detection 2$\sigma$ upper limit of $\le 5.8\times 10^{41}\ \rm{erg}\ \rm{s}^{-1}$.
Because of the low background and high sensitivity of our HST/WFC3 observations, 
this 2$\sigma$ upper limit is one order of magnitude deeper than the groundbased result in {\citet{nagao05}} for galaxies at $z\sim6$. 
%The rest-frame equivalent width of \ion{He}{2} is $0.0\pm 3.5$\AA. 

Using the HST $J-$ and $H-$band photometry, we show that BDF-521 has a blue rest-frame UV continuum, with a spectral slope $\beta=-2.32\pm0.43$ (Eq. 1). Early results from the HUDF09 team suggested that the high-redshift galaxy candidates at $z\gtrsim7$ have extremely blue UV slopes of $\beta\sim -2.0\pm 0.2$ for $L^*_{z=3}$ galaxies \citep{bouwens10, oesch10}.
%, and $\beta \approx$ $-3$ for 0.1 $L^*_{z=3}$ galaxies (Bouwens et al. 2010). %However, several work argue against these extreme blue slopes (e.g., Finkelstein et al. 2010, 2012; %McLure et al. 2011; Dunlop et al. 2012a; 
%Bouwens et al. 2012a). 
 The UDF12 campaign provided significantly deeper F160W imaging data for the spectral slope determination %(e.g., $\beta= 4.43\times(J125-H160 )-2) $ 
at redshifts $z\sim7-8$, added F140W imaging data to reduce potential observational biases 
(Robertson et al. 2013). 
This work measures the spectral slope of $\beta=-1.8\pm0.1$ for galaxies with luminosities of -20 $<$ M$_{\rm{UV}}$ $<$ -19 at $z\sim 7-8$. 
BDF-521 has an absolute magnitude $M_{\rm{AB}} (1500 \rm{\AA}) = -19.8$, corresponding to $0.4\times L^*_{z=3}$, and comparable with bright galaxy candidates in UDF12 campaign. The slope of BDF-521 is $\beta=-2.32\pm0.43$, consistent to within 1-$\sigma$ with UDF12 results as well as HDF09 results.
%\citet{bouwens10} studied the value of the UV-continuum slope $\beta$ in the Hubble Ultra Deep Field (HUDF). For luminous $L^*_{z=3}$ galaxies \citep{steidel99}, $\beta\sim -2.0\pm 0.2$. For lower luminosity 0.1$L^*_{z=3}$ galaxies, $\beta \sim -3.0\pm 0.2$.

\begin{table} [!h] 
\caption{HST Photometry of galaxy BDF-521} 
\label{table:F130N_S}	
% is used to refer this table in the text
\centering 
\begin{tabular}{c c c c c c} 
\hline\hline 
Band & Aperture\footnote{The circle aperture for photometry is 0.48$''$ for all the three different bands.}  & Flux(erg s $^{-1}$ cm$^{-2}$ \AA$^{-1}$) & m$_{\rm{AB}}$ & M$_{\rm{AB}}$\\	
% table heading
\hline 
F125W & 0.48'' &  $10.5\pm0.47\times10^{-21}$ & $27.06\pm0.05$ & $-19.93\pm0.05$\\
\hline
F160W & 0.48'' &  $6.48\pm0.38\times10^{-21}$ & $27.13\pm 0.06$ & $-19.86\pm0.06$ \\
\hline
F132N & 0.48'' &  $6.88\pm1.48\times10^{-21}$ & $27.43\pm0.22$ & $-19.56\pm0.22$\\
\hline
\end{tabular}
\end{table}

Fig.1 shows that BDF-521 is an extremely compact galaxy. Using the F125W broadband image, we measured its size and morphology based on Galfit measurements (Peng et al. 2002) and corrected for WFC3/F125W PSF broadening. 
The half-light radius of BDF-521 is 0.12'' $\pm$ 0.01'' in F125W. This corresponds to $0.64\pm0.06$ kpc (Fig. 1) and 
 is consistent with those of $L^*_{z=3}$ $z_{850}$-dropout  galaxy candidates at z $\sim 7$ detected in the Hubble Deep Field \citep{ono13}.

\section{Discussions}

We do not detect the He II emission line from BDF-521 at $z= 7.01$. However, we put the strongest upper limit on the SFR$_{\rm{PopIII}}$ to date. Combined with our He II upper limit on galaxies IOK-1 \citep{cai11}, we could constrain the formation of Pop III stars in two highest-redshift spectroscopically-confirmed galaxies at $z\sim7$. Note that 
He II emission, although at lower levels than for Pop III stars, can be arisen from other sources, such as photoionization by AGN, %stellar winds from Wolf-Rayet stars, 
or metal-poor nebular emission around massive stars. Due to the non-detection results in this paper, we assume Pop III stars are the only source of He II emission, and the upper limit of metal-free stars derived under this assumption is a secure limit.

Under the ``standard" assumptions \citep{schaerer02, schaerer03, raiter10}$-$ ionization bounded nebula, constant electron temperature and density, and Case B recombination $-$ the He recombination lines are fully specified and their luminosity is propotional to the ionizing photon production rate Q in the appropriate  
energy range. Following \citet{raiter10}, the relation between line luminosity and He II ionizing photon production rate $Q$(He$^+$) can be expressed as: 

\begin{equation}
L_{\rm{B}} (\rm{He} II \lambda1640)= \it{Q}(\rm{He^+}) \times c,
\end{equation}
where $c= 5.67 \times 10^{-12}$ erg for Z $<$ 1/50 Z$_\odot$ and 6.04$\times 10^{-12}$ erg for Z $>1/50$ Z$_\odot$ \citep{raiter10}. Eq. (3) can also be expressed by the following equation \citep{schaerer02}: 
\begin{equation}
L_{\rm{B}} ({\rm{He} II \lambda1640})=L_{\rm{1640, norm}} (\frac{\rm{SFR_{PopIII}}}{\rm{M_{\odot}\ yr^{-1}}}). 
\end{equation}
Under the Case B recombation model, the He ionizing photon production rate $Q(\rm{He^+}$) for a given SFR$_{\rm{PopIII}}$ mainly depends on two parameters according to theoretical investigations \cite[e.g.][]{schaerer02, raiter10, 
woods13}: (a) IMF, (b) mass loss. Therefore, given Eqs. (4) and (5), $L_{\rm{1640, norm}}$ also depends on these two parameters.  %Top-heavy IMF, strong mass loss,  will produce higher photons production rate (higher $Q(\rm{He})$ value) for a given SFR$_{\rm{PopIII}}$. Strong mass loss would p  

A top-heavy IMF will yield a higher photon production rate (higher $Q(\rm{He^+})$ value) for a given SFR$_{\rm{PopIII}}$. 
When the metallicity is smaller than the critical metallicity, with $Z_{\rm{crit}}\sim 10^{-4}\ Z_\odot$, 
stars are formed predominantly massive (Bromm et al. 2001), and the IMF 
should be much more top heavy than a normal IMF (Bromm et al. 2001). A few theoretical studies 
suggested that, at $z\gtrsim 7$, a significant fraction of galaxies with gas masses $\lesssim 10^{8}$ M$_\odot$ 
contain near pristine material of $Z< Z_{\rm{crit}}$ (e.g., Maio et al. 2013). 
The classical picture of Pop III star formation holds that the Pop III IMF is extremely top-heavy compared with a normal galactic IMF, and the typical mass for Pop III stars could be very massive 
of $\sim$ 100 M$_\odot$ \cite[e.g.][]{abel02}. %However, a new picture has emerged during the past few years.
However, recent simulations suggest that dynamical effects can cause the protodisk to fragment into multiple clumps; which may result in the formation of binary or multiple systems \citep{turk09, stacy10, clark11, greif12, dopcke13}. Using a grid code to study fragmentation with the highest resolution per Jeans length in minihalos, \citet{latif13} argue that a protostar of $\ge$ 10 M$_\odot$ can form via turbulent accretion. Thus, Pop III stars could have a lower characteristic mass than the previously suggested  $\sim 100$ M$_\odot$. Our non-detections of strong He II $\lambda1640$ for two galaxies  at $z=7$  disfavors the models of very massive Pop III stars dominating bright galaxies with $L\gtrsim0.5\ L^*_{z=3}$ at $z\sim7$.  
Table 2 demonstrates different upper limits of SFR$_{\rm{PopIII}}$ due to different mass ranges. A less top-heavy IMF corresponds to a weaker He II $\lambda1640$ line for a given of SFR$_{\rm{PopIII}}$. 

%For the shape of the IMF, \citet{dopcke13} suggest that, although the fragmentation could still happen in metal-free clouds, the matter spectrum flattens at $Z< 10^{-5}Z_{\odot}$, increasing the relative importance of the high-mass end and making the IMF top-heavy compared with a normal galactic IMF. Both a Salpeter IMF \citep{schaerer02, scannapieco03} and a log-normal IMF \citep{tumlinson06} have been used in model calculations. Table 2 shows a factor of six difference in inferred SFR arising from different mass cutoffs. The less top-heavy IMF corresponds to a weaker He II $\lambda1640$ line for a given of SFR$_{\rm{PopIII}}$. 

% and such less top-heavy IMFs make it more difficult to use He II line to constrain $f_{\rm{III}}$ at high redshift with the current facilities. 

Pop III stars with strong mass loss would yield a higher He II $\lambda1640$ luminosity for a given SFR$_{\rm{PopIII}}$. 
%For massive stars at very low metallicity, Heger et al. (2003) suggests that the line-driven winds are very low, and mass loss of Pop III stars is negligible.  
%However, recent observations support that mass loss in luminous blue variables (LBV) are much more important than was 
%previously estimated, and 
%strong LBV mass losses are seen in low metallicity dwarf galaxies (Smith 2014). Also, a few Type IIn supernovae and super-luminous supernovae occur in low metallicity dwarf galaxies (Smith 2014; Neill et al. 2011; Stoll et al. 2011), supporting that the low metallicity 
%could exhibit strong eruptive mass loss.
A commonly used approach to estimate stellar mass loss at low metallicity is to extrapolate the metal dependent mass loss to very low metallicity, even to metallicity lower than $Z_{\rm{crit}}$. Recent observations support that mass loss in luminous blue variables (LBV) are important, and
such strong LBV mass losses are seen in low metallicity dwarf galaxies, and are relatively 
insensitive to metallicity (Smith 2014, Neill et al. 2011). Thus, it is not safe to assume that massive stars at
very low metallicity suffer much less mass loss on average.  
From Table 2, given upper limit of He II emission, stronger mass loss results in a more stringent upper limit on SFR$_{\rm{PopIII}}$.

%The exact conversion from He II luminosity to Pop III SFR also depends on the model of 
%photoionization. \citet{schaerer02} calculates nebular emission lines using standard case B recombination and the 
%constant ionizing parameter. Eq. (4) is derived under these assumptions.  
%However, for low metallicity nebulae ionized by Pop III stars, the case B predictions  may have non-negligible deviations from real nebular astrophysics \citep{raiter10}. In that scenario, the the SFR$_{\rm{PopIII}}$ upper limits derived from Eq. (4) need to be qualitatively revised here. 
%According to \citet{raiter10}, for a given He II luminosity, SFR$_{\rm{PopIII}}$ could be a few times higher, depending on the ionizing parameters, e.g., the He number density $n(\rm{He})$, inner radius of the nebula $r_{in}$, and He ionizing photon production rate $Q(\rm{He^+})$. 

For BDF-521, previous ground-based observations suggest 
a extreme blue slope of $\beta =-4$ \citep{vanzella11}. Our HST measurement of the 
rest-frame UV luminosity has high S/N in both $J$ and $H$ bands. Based on that photometry, we get a normal slope of $\beta=-2.3$ (Eq. 2). 
We derive the UV-based unobscured SFR from the rest-frame UV continuum. 
The  UV-based overall SFR of BDF-521 is $\sim 5 \pm 0.2$ M$_\odot$ yr$^{-1}$, assuming the conversion for a 
Salpeter IMF between 0.1 to 100 M$_\odot$ (Madau et al. 1998). Note that the conversion of the UV-based overall 
SFR assumes a normal stellar population, without the correction for Pop III stars with a top-heavy IMF. This conversion 
is reasonable given the non-detection of strong He II emission. Normally, one expects that a broad red wing of the Gunn-Peterson (GP) trough would suppress the Ly$\alpha$ lines. As such, the Ly$\alpha$-based SFR may also be an underestimate. However, for BDF-521, the Ly$\alpha$-based SFR (\S2) is close to 
UV-derived SFR. This measurement suggests little extinction of  Ly$\alpha$ relative to the continuum, and supports that BDF-521 has ionized a significant volume around itself, and the broad GP trough does not reach Ly$\alpha$ emission.

If we assume the Pop III stars have an IMF with 50$\lesssim$ M/M$_\odot$ $\lesssim$500 and no mass loss, 
the ratio of SFR$_{\rm{PopIII}}$ to the overall SFR ($f_{\rm{III}}$) in BDF-521 is $<$ 20\%. If we assume Pop III stars have a Salpeter IMF with 50$\lesssim$M/M$_\odot \lesssim$ 1000 and strong mass loss, the ratio is $\lesssim$4\% (Table 2). 
 Kulkarni et al. (2013) use a semi-analytic model of galaxy formation  to argue that measurements of relative abundances in high-z damped Ly$\alpha$ systems (DLAs) can place constraints on the Pop III IMF. They show that the fractional contribution of very high-mass Population III stars to the ionization rate must be $z<$10\% at  $z=10$. Our observations are consistent that constraint. Combined with our 
previous upper limit from the galaxy IOK-1 (Cai et al. 2011), our observations targeting He II emission suggest that Pop III star formation with a top-heavy IMF could not significantly contribute to the overall star formation in a
$L \sim L^*(z=3)$ galaxy at $z\sim 7$. 
%We convert the rest-frame UV luminosity to an unobscured total SFR. 
%The conversion of UV luminosity to an overall SFR assumes a conversion factor
%between UV luminosity and SFR for normal Pop I and Pop II stellar populations, without correcting
%for Pop III stars with a top-heavy IMF. 
%The rest-frame UV luminosity %flux density is 10.5 $\pm$ 0.5 $\times$ 10$^{-21}$ ergs s$^{-1}$ cm$^{-2}$ \AA$^{-1}$, which 
%corresponds to an overall SFR of $\sim$ 
%5$\pm$0.2 M$_\odot$ yr$^{-1}$ (Madau 1998). 
%(This result could change by a factor of 2 given the systematic uncertainties associated with the IMF and metallicity assumptions). 
%Using the fiducial top-heavy Pop III Salpeter IMF of 50 - 1000
%M$_\odot$ and mass loss, we derived a $2\sigma$ upper limit of  M$_\odot$ yr$^{-1}$ 
%for SFR$_{\rm{PopIII}}$.
%corresponds to an upper limit of 6.0 $\times$ 10$^{-21}$ ergs s$^{-1}$ cm$^{-2}$ \AA$^{-1}$ in F125W flux density. 
 %Thus, BDF-521 is not dominated by very massive Pop III stars. 

If Population III stars are not from very top-heavy IMFs (see e.g., Stacy et al. 2010), 
the He II emission of galaxies at $z>6$ is too weak for current telescopes to detect with a reasonable exposure time. Also, the clear, 
 solid signatures of a Pop-III dominated galaxy might require reaching the continuum level for fainter high-z galaxies. 
\citet{zackrisson12} suggest that the Pop III galaxies can experience siginificant Lyman continuum leakage (also, see \citealp{inoue11}), %to render the nebular contributions to their SEDs negligible , 
and that such objects would stand out in surveys by the 
{\it James Webb Space Telescope (JWST)} due to their extremely blue rest-frame FUV continuum slopes. 
 \citet{mitra11} and \citet{choudhury07} predict that if the observations reach a magnitude of $J_{110, AB}\sim 31-32$, $\sim 15-30\   \rm{arcmin}^{-2}$, Pop III-dominated sources residing halomass of $M< 10^{9} M_\odot$ will be detected at $z>7$.
Future facilities, especially {\it JWST} %and 30-m Giant Segmented Mirror Telescopes, 
are expected to probe both the continuum and recombination lines 
for faint galaxies with low halo masses ($M\sim 10^9 M_\odot$) at $z\gtrsim10$ with reasonable exposure times, and, in doing so, to unveil the signatures of the earliest star formation 
in the universe.

%There is no quantitative formula
%for He II flux calculations for different photoionization models.
%t is because the ionization parameter does not need to be constant
%in the real nebulae, where the gas might be a mixture of regions of
%different densities, the geometry might be complicated which all affect the ionization
%paratemer. However, for low metallicity nebulae ionized by very hot Pop III stars, the case B predictions for line and continuum emission may have non-negligible deviations from real nebular astrophysics (Raiter et al. 2010). Therefore, the SFR$_{\rm{PopIII}}$ upper limits derived from Eq. (5) then need to be qualitatively revised here. According to Raiter et al. (2010), for a given He II lumi- nosity, SFRPopIII could be a few times higher, depending on the ionizing parameters, e.g., the hydrogen number density $n(\rm{H})$, inner radius of the nebula $r_{in}$, and hydrogen ionizing photon flux $Q(\rm{H})$.

\begin{table} [!h]
\caption{Upper Limit of Pop III Star Formation Rate of BDF-521 at $z=7.01$} 
\label{table:PopIII_SFR}	
% is used to refer this table in the text
\centering 
\begin{tabular}{|c | c | c| c | c |} 
\hline\hline 
IMF(Salpeter) & Mass Loss & $L_{\rm{norm}, 1640} $ & 2-$\sigma$ upper limit of $\rm{SFR}_{\rm{Pop} III} $ & 2-$\sigma$ upper limit of $\rm{f_{III}}$\footnotemark[1]\\
                    &                & ($\mbox{ergs}\ \rm{s}^{-1}$) & $ (\rm{M}_{\odot} \rm{yr}^{-1})$ &\\
% table heading
\hline
$1\lesssim M/M_{\odot}\lesssim500$ & No & $ 9.66\times10^{40}$ & $6.0$ & 100\%\\
\hline 
$50\lesssim M/M_{\odot}\lesssim500$ & No & $6.01\times10^{41}$ &  1.0 & 20\%\\
\hline
$1\lesssim M/M_{\odot}\lesssim500$ & Yes & $3.12\times10^{41} $ & $1.9$& 38\%\\
\hline
$50\lesssim M/M_{\odot}\lesssim1000$ & Yes & $2.33\times10^{42}$ & 0.2 & 4\%\\
\hline
\end{tabular}
\end{table}
{\small Note: $f_{\rm{III}}$ represents the ratio of the Pop III star formation rate to the overall unobscured star formation rate.}

%Observationally, Scannapieco et al. (2006) suggests a lower mass cutoff of 0.8M$_\odot$ based on metal abundances in Galactic halo stars. 

\section{Acknowledgement}
 We especially thank the anonymous referee for insightful comments which have significantly improved the Letter. 
%We thank Stefano Casertano, Daniel Schaerer, and Haojing Yan for useful discussions.  
Support for this work was provided by NASA through grant HST-GO-12487  from the Space Telescope Science Institute, which is operated by AURA, Inc., under NASA contract NAS5-26555. 
   
%
% Subsection heading
%

%
% References
%
\paragraph{}

%\end{document}

\clearpage
\setcounter{figure}{0}

\figurenum{1}
\begin{figure}[tbp]
\epsscale{1}
\label{fig:images}
\plotone{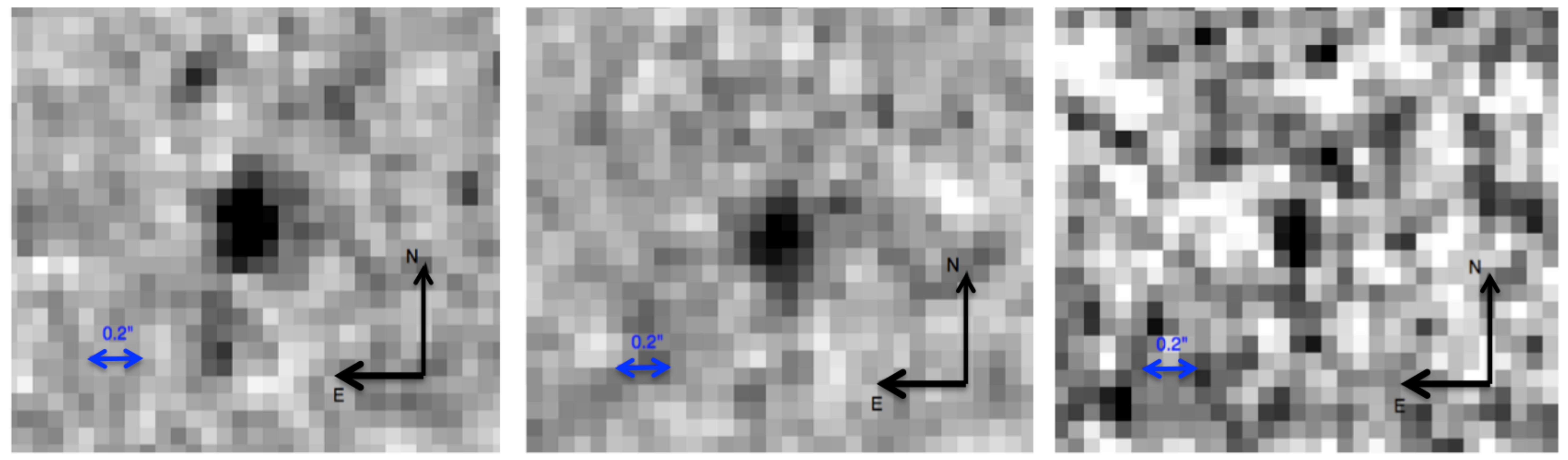}
\caption{High resolution images of the galaxy BDF-521 in the F125W (left), F160W (middle) broadband filters and the F132N (right)
narrowband filter. %with contours spaced by 1.4 sky rms for the F125W and F160W images and 1 sky rms for the
%F132N image. %The black elliptical aperture is determined by the F130N image. 
BDF-521 has a luminosity of $0.4\times L^*_{z=3}(\lambda=1500\ \rm{\AA})$ at $z=7.01$, and the half light radius is about $0.64\pm0.06$ kpc. 
The BDF-521 is a compact galaxy.  The size of BDF-521 is consistent with the size of $L^*_{z=3}(\lambda=1500\ \rm{\AA})$ $z_{850}$-dropout and $Y_{105}$-dropout galaxy candidates at z $\sim 7-8$ detected in the HUDF12 (Ono et al. 2013).  Photometric analysis of the F132N image do not reveal a strong He II $\lambda1640$ emission. }
\end{figure}

\figurenum{2}
\begin{figure}[tbp]
\epsscale{1}
\label{fig:testschemes}
\plotone{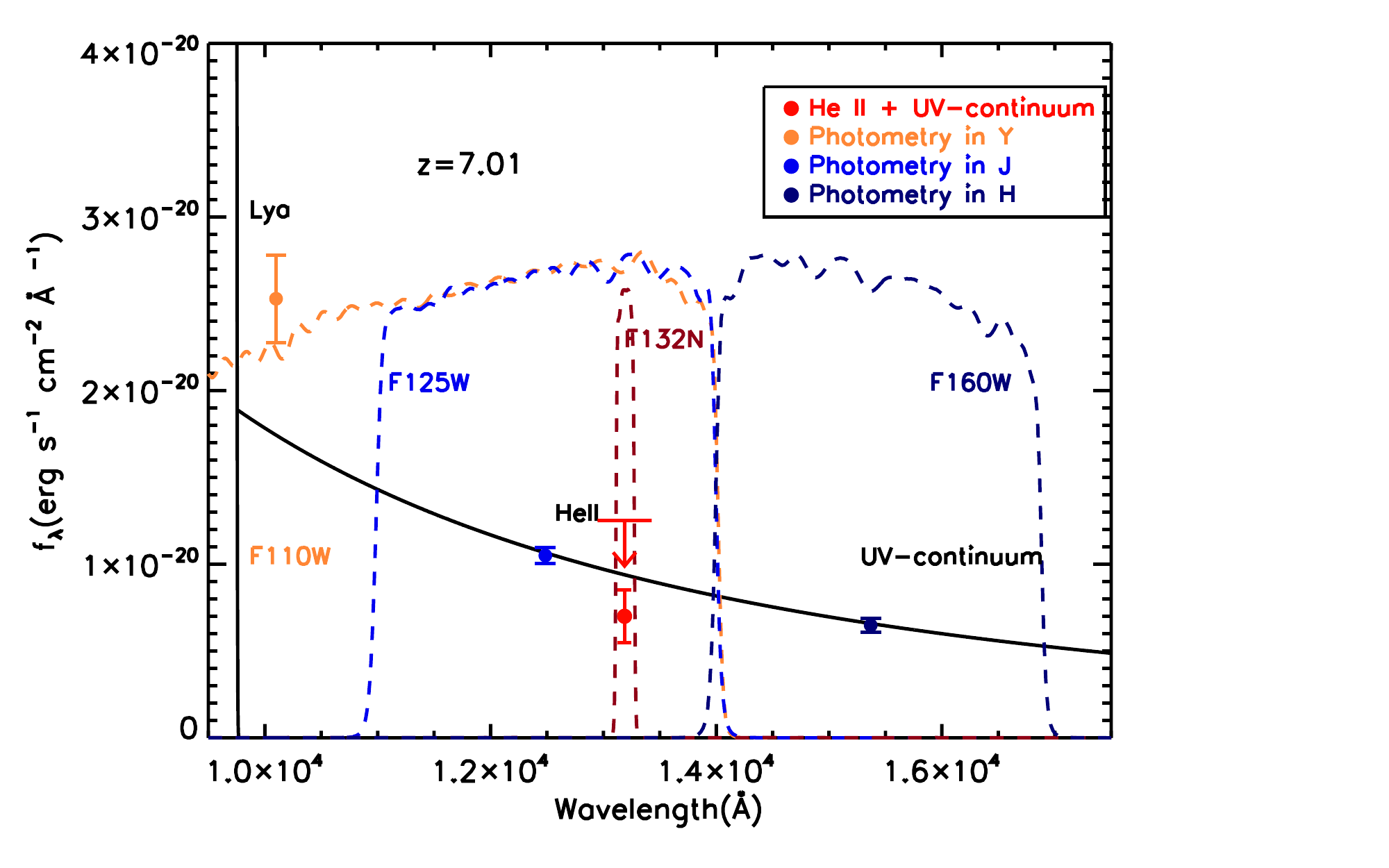}
\caption{Best-fit spectrum (black line) of the galaxy BDF-521 with the total UV continuum, as well as the Ly$\alpha$ (Vanzella et al. 2011) and He II lines (this work). The UV continuum is fit using our high S/N photometry in F125W (blue point with error bar), F132N (red point with error bar), and F160W (dark blue point with error bar) bands using Hubble Space Telescope (HST) (see Table 2). The filter response curves of WFC3/F110W (yellow dashed line), WFC3/F125W (blue dashed line), WFC3/F160W (dark blue dashed line), and WFC3/F132N (brown dashed line) are plotted.  In addition, photometry in four different bands is overplotted at the effective wavelength of each filter, where $Y$-band photometry (yellow point with error bar) is from Vanzella et al. (2011); and $J$-, $H$-, and narrowband photometry are from our HST observations. The red arrow presents the non-detection 2$\sigma$ upper limit of the He II $\lambda1640$ emission line. }
\end{figure}

%%%%%%%%%%%%%%% Table 2 %%%%%%%%%%%%%%%%%%%%%%%%%%%%%%%
%\begin{deluxetable}{ccccc}
%\centering
%\tabletypesize{\tiny}
%\topmargin 0.0cm \evensidemargin = 0mm
%\oddsidemargin=0mm
%\rotate
%\tablecaption{Pop III star formation rate}
%\tablehead{
%\colhead{IMF(Salpeter)} &
%\colhead{Mass Loss} &
%\colhead{$f_{1640} $} &
%\colhead{$\rm{SFR}_{\rm{Pop} III}} &
%\colhead{$\rm{f_{III}}$} \\
%\colhead{(1)}  & \colhead{(2)} & \colhead{(3)} & \colhead{(4)} & \colhead{(5)} 
%}
%\startdata
%$1\lesssim M/M_{\odot}\lesssim500$ & No &$ 9.66\times10^{40}$ & $6.8\pm4.7$& 34\%\\
%\hline 
%$50\lesssim M/M_{\odot}\lesssim500$ & No & $6.01\times10^{41}$ & $1.1\pm0.9$& 5.5\%\\
%\hline
%$1\lesssim M/M_{\odot}\lesssim500$ & Yes & $3.12\times10^{41} $ & $ 2.1\pm1.5$& 11\%\\
%\hline
%%$50\lesssim M/M_{\odot}\lesssim1000$ & Yes & $2.33\times10^{42}$ & $0.3\pm0.2$&1.5\%
%\enddata
%\tablecomments{\normalsize
%The contribution of Pop III stars to the observed SFR
%Column (1) official SDSS name;
%Column (2) luminosity of the power-law continuum at 3000\,\AA, $\lambda L_{\lambda}$(3000 \AA);
%Column (3) FWHM of broad Mg\,II\,$\lambda2800$, corrected for instrumental broadening;
%Column (4) flux of near-UV Fe\,II emission
%(integrated in the range of 2200--3090 \AA\ from the best-fit model);
%%Column (5) rest-frame equivalent width of UV Fe\,II emission.}
%\end{deluxetable}
\end{document}